\documentclass[10pt]{iopart}

\usepackage{graphicx}

\begin{document}

\title[Engineering the properties of the quinternary half-metallic Heusler
alloys]{Engineering the electronic, magnetic and gap-related
properties of the quinternary half-metallic Heusler alloys}

\author{K \"Ozdo\~gan\dag, E \c Sa\c s\i o\~glu\ddag, and  I Galanakis\S}

\address{\dag\ Department of Physics, Gebze Institute of Technology,
Gebze, 41400, Kocaeli, Turkey }

\address{\ddag\ Institut f\"ur Festk\"orperforschung, Forschungszentrum
J\"ulich, D-52425 J\"ulich, Germany and Fatih University, Physics
Department, 34500, B\" uy\" uk\c cekmece,  \.{I}stanbul, Turkey}

\address{\S\ Department of Materials Science, School of Natural
  Sciences, University of Patras,  GR-26504 Patra, Greece}

\ead{kozdogan@gyte.edu.tr,e.sasioglu@fz-juelich.de,galanakis@upatras.gr}

\begin{abstract}
We review the electronic and magnetic properties of the
quinternary full Heusler alloys of the type
Co$_2$[Cr$_{1-x}$Mn$_x$][Al$_{1-y}$Si$_y$] employing  three
different approaches : (i) the coherent potential approximation
(CPA), (ii) the virtual crystal approximation (VCA), and (iii)
supercell calculations (SC). All three methods give similar
results and the local environment manifested itself only for small
details of the density of states. All alloys under study are shown
to be half-metals and their total spin moments follow the
so-called Slater-Pauling behavior of the ideal half-metallic
systems. We especially concentrate on the properties related to
the minority-spin band-gap. We present the possibility to engineer
the properties of these alloys by changing the relative
concentrations of the low-valent transition metal and $sp$ atoms
in a continuous way. Our results show that for realistic
applications, ideal are the compounds rich in Si and Cr since they
combine large energy gaps (around 0.6 eV), robust half-metallicity
with respect to defects (the Fermi level is located near the
middle of the gap) and high values of the majority-spin density of
states around the Fermi level which are needed for large values of
the perfectly spin-polarized current in spintronic devices like
spin-valves or magnetic tunnel junctions.
\end{abstract}


\submitto{\JPD}

\maketitle

\twocolumn

\section{Introduction\label{sec1}}

The rapid growth of the field of magnetoelectronics, also known as
spintronics, brought to the attention of scientists new phenomena
\cite{Zutic,Felser,Zabel}. One of the most interesting concepts in
spintronics is the half-metallicity
\cite{book,Review1,Review2,Westerholt,Reiss,Sakuraba,Dong}.
Half-metals are hybrids between normal metals and semiconductors.
The majority-spin band is crossed by the Fermi level as in a
normal metal while the Fermi level falls within a gap in the
minority-spin band as in semiconductors leading to a perfect
100\%\ spin-polarization at the Fermi level \cite{Review1} Such
compounds should have a fully spin-polarized current and be ideal
spin injectors into a semiconductor, thus maximizing the
efficiency of spintronic devices \cite{Wolf}.  de Groot and his
collaborators in 1983 were the first to predict the existence of
half-metallicity in the case of the intermetallic semi-Heusler
alloy NiMnSb \cite{deGroot} and the origin of the gap seems to be
well understood \cite{GalaHalf,gap}. The half-metallic character
of NiMnSb in single crystals seems to have been well-established
experimentally. Infrared absorption \cite{Kirillova95} and
spin-polarized positron-annihilation \cite{Hanssen90} gave a
spin-polarization of $\sim$100\% at the Fermi level.
First-principles electronic structure calculations were successful
in explaining the origin of half-metallicity in Heusler alloys in
terms of the hybridization between the $d$-orbitals of the
transition metal atoms and as it was shown the half-metallicity is
also closely related to the total spin magnetic moment in the unit
cell \cite{GalaHalf}. However, in many other experiments on
low-dimensional structures the half-metallicity has not been found
since the properties which have been measured were surface and
interface sensitive.

Although semi-Heuslers have initially monopolized the interest,
the last years the interest has been shifted to the so-called
full-Heusler compounds like Co$_2$MnAl. Webster was the first to
synthesize such alloys already in 1971 \cite{Webster} and almost
20 years later it was  argued in two papers by a japanese group
that they should be half-metals \cite{Ishida-Fujii}. The emergence
of the field of magnetoelectronics brought again at the center of
scientific interest these alloys \cite{Picozzi,GalaFull} and
first-principles calculations have explained the origin of
half-metallicity, demonstrating also that the perfect
half-metallic full Heusler alloys show the so-called
Slater-Pauling behavior; the total spin moments in the unit cell,
$M_t$ in $\mu_B$, equals the total number of valence electrons in
the unit cell minus 24 since there are exactly 12 occupied
minority-spin states \cite{GalaFull}. This rule is of interest
since it provides an easy way to connect half-metallicity with the
total spin-moment in the unit cell which is easily determined
using various experiments like SQUID measurements. Although
several full-Heusler alloys have been shown to be half-metallic in
their bulk form \cite{GalaFull,Podlucky}, ab-initio calculations
have shown that the surfaces \cite{Gala-Surf} and interfaces of
the full-Heusler compounds \cite{Interfaces} loose their
half-metallicity (Hashemifar and collaborators have shown that it
is possible to restore half-metallicity at some surfaces
\cite{Hashemifar}). Except interface states also temperature
driven excitations \cite{Chioncel,MavropTemp,Skomski} and defects
\cite{Picozzi04,Orgassa} seem to destroy half-metallicity. Also
some other important aspect of these alloys like the orbital
magnetism \cite{orbit}, the doping \cite{APL}, the structural
stability \cite{Block}, the appearance of ferrimagnetism
\cite{PSS-RRL,SSC,Mn2VZ} and antiferromagnetism \cite{AFM} and the
interplay of exchange interactions \cite{Sasioglu,Kurtulus} have
been addressed in literature.

\begin{figure*}
\centering
\includegraphics[scale=0.25]{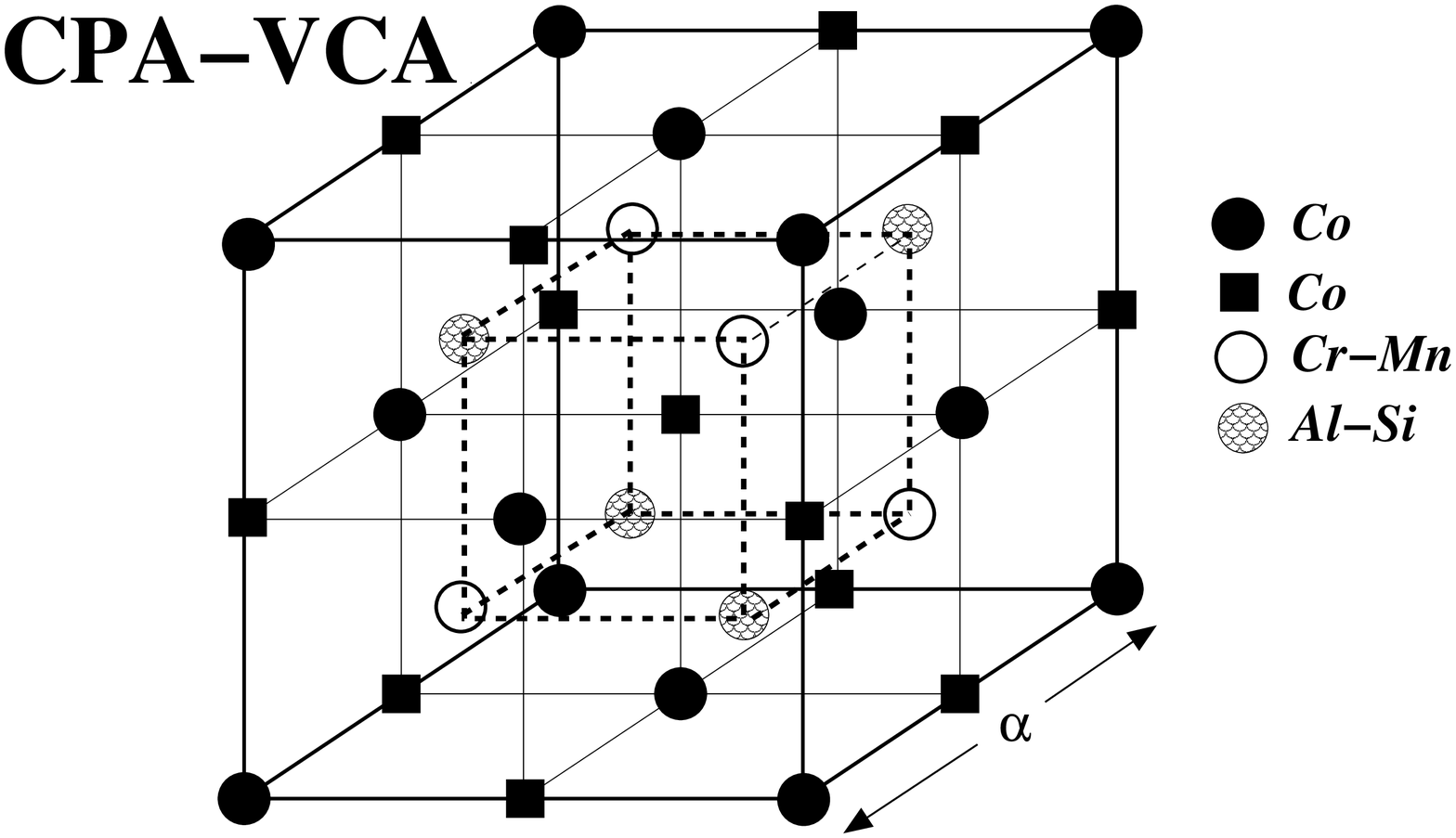}
\includegraphics[scale=0.25]{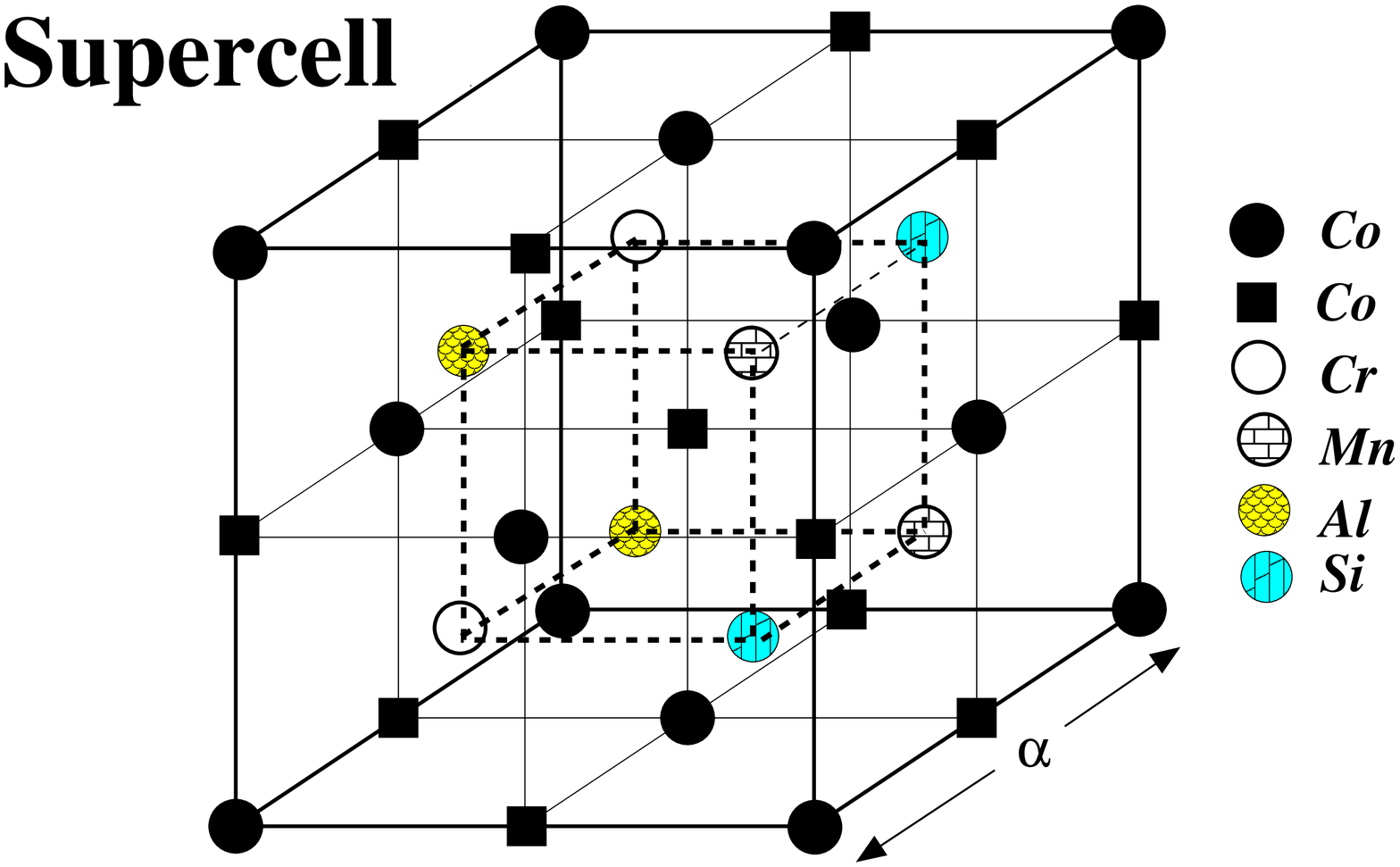}
\caption{Schematic representation of the structure used for the
Co$_2$[Cr$_{0.5}$Mn$_{0.5}$][Al$_{0.5}$Si$_{0.5}$] alloy. On the
left the structure used for the CPA and VCA calculations. In CPA
the Cr-Mn site is occupied by Cr atoms with a probability of 50\%\
and by Mn atoms with a probability of 50\%\ and the Al-Si site is
also occupied by both Al and Si atoms with 50\%\ probability for
each one. In VCA, the  Cr-Mn site is occupied by a pseudoatom with
a fractional number of valence electrons (24.5 electrons) and the
Al-Si site by a pseudoatom with 13.5 valence electrons. The unit
cell for both CPA and VCA calculations contains four sites. On the
right panel we present the structure for the supercell
calculations where we take a double unit cell with respect to VCA
and CPA containing eight atoms (Co$_4$CrMnAlSi compound). In the
case where $x$ and/or $y$ take 0.25 or 0.75 as a value we have a
unit cell with 16 atoms for the supercell calculations. Note also
that there are two inequivalent Co sites in all cases which have
the same environment rotated by $\pi/2$.} \label{fig1}
\end{figure*}

Over the last two years the interest in full-Heusler alloys
containing cobalt has been focused on the so-called quaternary
Heusler alloys, which are found to present half-metallicity as
long as the corresponding perfect parent compounds are half-metals
\cite{GalaQuat,GalaJAP}. Several authors studied the properties of
Co$_2$[Cr$_{1-x}$Fe$_{x}$]Al as a function of the concentration
$x$ \cite{Antonov,Miura}  as well as the
Co$_2$[Mn$_{1-x}$Fe$_{x}$]Si alloys \cite{Balke} due to the fact
that ab-initio calculations, including the on-site Coulomb
repulsion (the so-called Hubbard $U$), have shown that Co$_2$FeSi
can reach a total spin magnetic moment of 6 $\mu_B$ which is the
largest known spin moment for a half-metal \cite{Kandpal,Wurmehl}.

\section{Description of present calculations\label{sec2}}

In a recent paper \cite{GalaJAP} we employed the full--potential
nonorthogonal local--orbital minimum--basis band structure scheme
(FPLO) \cite{koepernik1,koepernik2} to study the properties of the
quaternary Heusler compounds  Co$_2$[Cr$_{1-x}$Mn$_x$]Z,
Co$_2$[Cr$_{1-x}$Fe$_x$]Z and Co$_2$[Mn$_{1-x}$Fe$_x$]Z where Z
stands for Al, Ga, Si, Ge or Sn. In the case of
Co$_2$[Cr$_{1-x}$Mn$_x$]Z compounds it was shown by standard
electronic structure calculations employing the local-spin-density
approximation (LSDA) \cite{LDA} that the compounds are
half-metallic for all concentrations. This yielded the idea that
if one studied the case of quinternary alloys, allowing for an
additional degree of freedom since also the chemical elements at
the Z site are mixed, one can engineer in a continuous way the
properties of these alloys related with the gap, thus the width of
the minority-spin band gap, the position of the Fermi level and
the majority-spin density of states at the Fermi level. Thus we
decided to study the family of
Co$_2$[Cr$_{1-x}$Mn$_x$][Al$_{1-y}$Si$_y$] compounds since all
four parent compounds are half-metals and such compounds seem
feasible experimentally. We have used the experimental lattice
constants for the perfect compounds containing Mn
\cite{lattice,lattice2}, 0.5756 nm for Co$_2$MnAl and 0.565 nm for
Co$_2$MnSi, and considered that the lattice constant varies
linearly with the concentration $y$ of the Al and Si atoms. We
assumed that the substitution of Cr for Mn does not change the
lattice constant since no evidence is known on the exact behavior
of the lattice. This tactic is different than the one used in
reference \cite{GalaQuat} where it was assumed that the lattice
constant varies linearly with the concentration $x$ of the
transition-metal atom but both methods give lattice constants
within less than 1\% difference and thus practically identical
results.

To perform our calculations we have employed three different
formalisms. Firstly the coherent-potential approximation (CPA)
initially developed by Blackman and collaborators in 1971
\cite{CPA}. In 1997 Koepernik et al have extended this formalism
and have implemented it in a Linear Combination of Atomic Orbitals
(LCAO) method \cite{koepernik1} and later they expanded it to to
incorporate it in the FPLO method \cite{koepernik2}. This was made
possible due to the localized-character of the orbitals in this
method which allows for a single-site representation of the space.
In CPA each Y (or Z) site is occupied by both Cr and Mn (or Al and
Si) atoms with a probability given by the respective concentration
of each chemical type. Second, we used the so-called virtual
crystal approximation (VCA). In VCA for the Y (or Z) site we
substitute both Co and Mn atoms (or Al and Si atoms) with an atom
with fractional number of electrons $(1-x)*z^{Cr}+x*z^{Mn}$ (or
$(1-y)*z^{Al}+y*z^{Si}$) where $z^{Cr}$ the number of electrons of
Cr   and similarly for the other chemical elements. Obviously in
VCA only the total spin magnetic moment in the unit cell and the
total DOS have a physical meaning since we can not project the
properties of the pseudoatom on the different atoms. We represent
the structures within both CPA and VCA in the left part of figure
\ref{fig1}. The lattice is that of an fcc with four atoms as basis
set along the diagonal: two Co atoms, one Cr-Mn site and one Al-Si
site. Both CPA and VCA do not take into account the short-range
interactions since the real crystal is replaced by one where each
Y or Z site contains two different chemical type atoms. In reality
this is not true and we performed supercell (SC) calculations to
examine this effect. We present the structure in the right part of
figure \ref{fig1} in case that $x=0.5$ and $y=0.5$ where the unit
cell contains eight atoms; in reality the correct chemical formula
is Co$_2$CrMnAlSi. In case that at least one of $x$ or $y$ is 0.25
or 0.75 the unit cell contains 16 atoms. To compare between the
three different approaches when we present the total density of
states (DOS) or the total spin magnetic moments we always scale to
a formula unit of four atoms.

Such an investigation of the gap-related properties of the Heusler
alloys is of interest due to the large variety of applications of
these alloys. Several experiments have been devoted to the study
of the structural and magnetic properties of the quaternary
Heusler alloys \cite{Elmers} and such films have been incorporated
both in magnetic tunnel junctions \cite{Marukame} and spin-valves
\cite{Kelekar}. Quinternary half-metallic full-Heusler alloys will
provide an additional tool to tune the properties of these films
and enhance the performance of the devices based on such films.

\begin{figure}
\centering
\includegraphics[width=\linewidth]{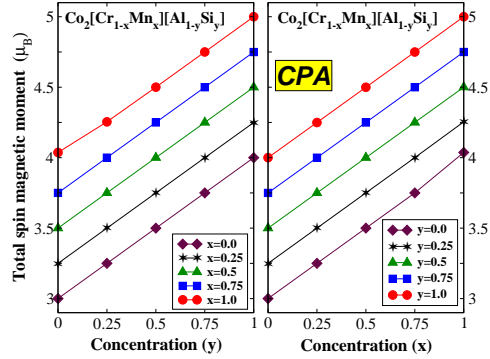}
\caption{Left panel : Total moment in $\mu_B$ as a function of the
concentration $y$ in Co$_2$[Cr$_{1-x}$Mn$_x$][Al$_{1-y}$Si$_y$]
using the CPA method. Different lines correspond to constant
values of $x$. Right panel : similar to the left panel as a
function of $x$ with different lines corresponding to constant
values of $y$. All compounds are half-metals showing the
Slater-Pauling behavior.} \label{fig2}
\end{figure}

\section{Total and atom-resolved spin magnetic moments \label{sec3}}

We will start our discussion on the quinternary alloys discussing
the magnetic spin moments. To present our results we discuss the
trends as a function of the concentration $x$ of the
transition-metal atoms Cr and Mn keeping the concentration $y$ of
the sp atoms constant and afterwards as a function of $y$ keeping
$x$ constant. This presentation allows for an easier discussion of
the results. In figure \ref{fig2} we have plotted the total spin
moment in the unit cell in $\mu_B$ within the CPA approach. If the
alloys under study are half-metals they should follow the
Salter-Pauling behavior for the total spin moments, $M_t$ :
$M_t$=$Z_t$-24 \cite{GalaFull}. The total spin moment in $\mu_B$
is just the number of uncompensated spins and thus the number
``24" arises from the 12 occupied minority spin states (for
details see reference \cite{GalaFull}). $Z_t$ denotes the total
mean number of valence electron and is given by the expression
$Z_t=2*z^{Co}+(1-x)*z^{Cr}+x*z^{Mn}+(1-y)*z^{Al}+y*z^{Si}$ where
$z$ denotes the valence of each atom. As it is presented in figure
\ref{fig2} the values of the total spin moments are spotted on top
of straight lines both when we plot the total moment as a function
of $y$ (left panel) or as a function of $x$ (right panel). A close
look on the figure reveals that all these straight lines follow
the Slater Pauling behavior for the full-Heusler alloys.
Co$_2$CrAl has 27 electrons and an ideal total spin moment of 3
$\mu_B$ which coincides with the calculated one. As we substitute
25\%\ of the Cr atoms by Mn ($x$=0.25) we increase the total spin
moment by 0.25 $\mu_B$ each time reaching a value of 4 $\mu_B$ for
Co$_2$MnAl. Also at each step when we substitute 25\%\ of the Si
atoms for Al again we increase the number of valence electrons by
0.25 and thus the total spin moment increases also by 0.25 $\mu_B$
at each step reaching the 4 $\mu_B$ for Co$_2$CrSi and 5 $\mu_B$
for Co$_2$MnSi.

\begin{figure}
\centering
\includegraphics[width=\linewidth]{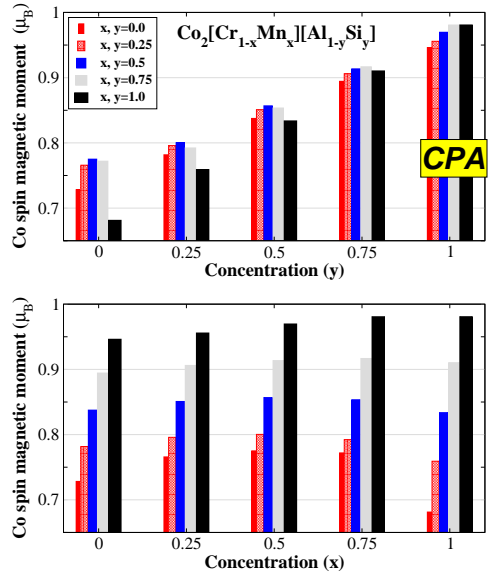}
\caption{Top panel : Co-resolved spin magnetic moment in $\mu_B$
as a function of the concentration $y$ in
Co$_2$[Cr$_{1-x}$Mn$_x$][Al$_{1-y}$Si$_y$] using the CPA method.
Different bars correspond to different values of $x$. Bottom panel
: similar to the top panel as a function of $x$ with different
bars corresponding to different  values of $y$.} \label{fig3}
\end{figure}

\begin{figure}
\centering
\includegraphics[width=\linewidth]{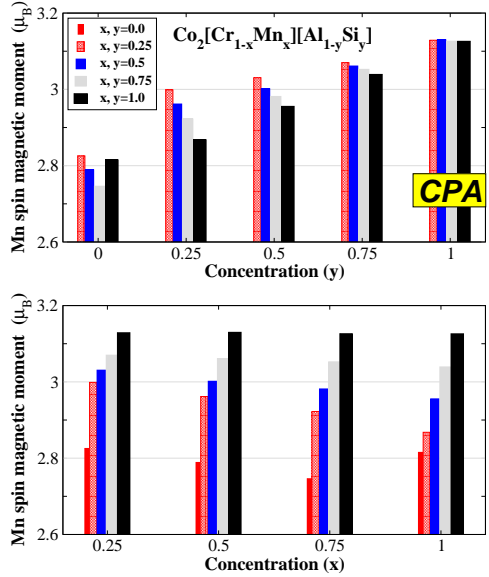}
\caption{Same as Fig. \ref{fig3} for the Mn atoms. The spin
magnetic moment has been scaled to one Mn atom.} \label{fig4}
\end{figure}

\begin{figure}
\centering
\includegraphics[width=\linewidth]{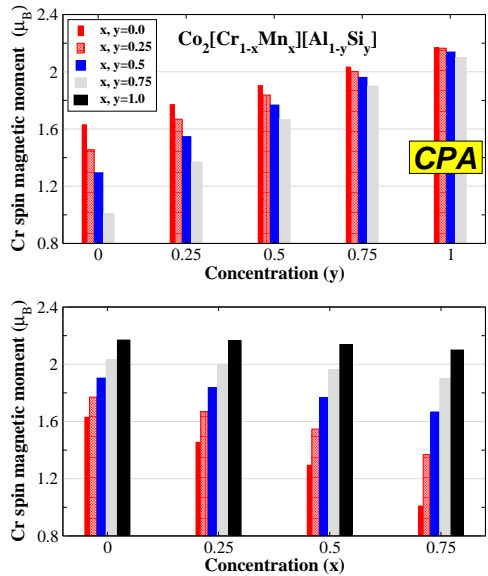}
\caption{Same as Fig. \ref{fig3} for the Cr atoms. The spin
magnetic moment has been scaled to one Cr atom.} \label{fig5}
\end{figure}

In figures \ref{fig3}, \ref{fig4} and \ref{fig5} we have plotted
the atom resolved spin moments for the transition metal atoms Co,
Cr and Mn, respectively, within the CPA approximation. In each
figure we present the moment as a function of the concentration
$y$ in the upper panel with different bars corresponding to
different $x$ and as a function of $x$ in the lower panel with
different bars corresponding to different values of $y$. We do not
present the spin moments of the $sp$ atoms since they are very
small and negative (around -0.1 $\mu_B$). The negative sign arises
from the $p$ states. The minority bonding $p$ states are
completely occupied while the majority bonding ones extend much
more in energy crossing the Fermi level resulting in a excess of
minority $p$ states with respect to the majority ones and thus a
negative spin moment. The behavior of the $sp$ atoms is
extensively discussed in references \cite{Review1} and
\cite{GalaFull} and is independent of the $x$ and $y$
concentrations.

For the Co spin moments in figure \ref{fig3} it is obvious from
the upper panel that the Co spin moment depends on the $y$
concentrations of the sp atoms. When $y=0$ and the compounds
contain only Al, the Co spin moments are quite small around
0.7-0.75 $\mu_B$. When 25\%\ of the Al atoms are substituted by
Si, the Co spin moments increase and reach the 0.8 $\mu_B$. When
$y=0.5$ the spin moments are around 0.85 $\mu_B$ and when $y=1.0$
and the compounds contain only Si, the Co spin moments are around
0.95 $\mu_B$. Thus in general as we increase the concentration of
the Si atoms by 25\%\ the Co spin moment increases by 0.05
$\mu_B$. Obviously the concentration of the Cr and Mn atoms does
not affect the Co spin moment as can be seen in figure \ref{fig3}.
The substitution of Cr by Mn does not change the hybridization of
the $d$-orbitals of the Co atoms with the $d$-orbitals of the
lower-valence transition metal atoms, Cr and Mn, and thus the Co
spin moment is not affected by the $x$ concentration. On the other
hand when we change the concentrations $1-y$ and $y$ of the Al and
Si atoms, the extra electronic charge occupies majority spin
electronic states of the transition metal atoms as in a rigid band
model and increases the spin moment of all transition-metal atoms
Co, Cr and Mn.

In figures \ref{fig4} and \ref{fig5} we present the spin moments
of the Cr and Mn atoms scaled to one atom. The conclusions drawn
for the Co spin magnetic moments hold generally also for the Mn
and Cr atoms although the fluctuations are more important than for
the Co atoms. Both Cr and Mn spin magnetic moments depend mainly
on the concentration $1-y$ and $y$ of the Al and Si atoms. Mn spin
moment starts from around 2.8 $\mu_B$ for $y=0$ corresponding to
the compounds containing only Al and goes up to 3.1 $\mu_B$ for
$y=1$ corresponding to the compounds where only Si is present.
Similarly the Cr spin moments reach a value of around 2.1 $\mu_B$
for the exclusively Si-based alloys ($y=1$). When $y=0$ or
$y=0.25$ the Cr spin moment present large fluctuations with
respect to the $x$ concentration being 1.6 $\mu_B$ for Co$_2$CrAl
and 0.9 $\mu_B$ for Co$_2$[Cr$_{0.75}$Mn$_{0.25}$]Al. This
behavior of the Cr spin moments has been explained in reference
\cite{PSS-RRL} where it was attributed to the pinning of the Fermi
level in a narrow band created by the Cr triple degenerated
majority $t_{2g}$ electrons which overlaps strongly with the
double-degenerated $e_g$ majority states, contrary to the behavior
of the Mn atoms \cite{SSC}. As shown in \cite{PSS-RRL} a small
shift of the Fermi level results in large changes of the Cr spin
moments. When we increase the concentration in Si atoms, the extra
electrons push the Cr majority states lower in energy and thus the
Cr spin magnetic moment is insensitive to the exact position of
the Fermi level and thus the Cr spin moment is the same for all
compounds with different $x$ value but the same $y$ value when
$y=$0.5, 0.75 or 1.0 (compounds rich in Si).

\begin{table*}
\centering \caption{Calculated spin magnetic moments in $\mu_B$
using a supercell construction  for the
Co$_2$[Cr$_{1-x}$Mn$_x$][Al$_{1-y}$Si$_y$] as a function of the
$x$ and $y$ concentrations. We do not present the spin moments of
the Al and Si atoms since they are very small (around -0.10
$\mu_B$ for all compounds). All compounds are half-metals and the
total spin moment is the ideal one predicted by the Slater-Pauling
rule (we have scaled the total spin moment to the elementary unit
cell containing four atoms only; the unit cell in the supercell
calculations contains either 8 or 16 atoms). In case that there
are more than one inequivalent atom of the same chemical kind, we
present the largest spin moment; the difference between this
moment and the other moments of the same chemical type are less
than 0.0.5 $\mu_B$ in all cases. In parenthesis the calculated
spin magnetic moments using the CPA approximation and in brackets
using the VCA approximation. Note that within VCA we have a
pseudoatom with 24.5 electrons instead of distinct Co and Mn atoms
and we include these spin moments both in Co and Mn columns.
\label{table1}}
\begin{tabular}{r|r|c|c|c|c}\hline\noalign{\smallskip}
$x$ & $y$ & $m^{Total}$  & $m^{Co}$ & $m^{Cr}$ & $m^{Mn}$   \\
& & & SC (CPA) [VCA] &  SC (CPA) [VCA]& SC (CPA) [VCA] \\ \hline
0.00 & 0.00  &  3.00 & 0.73 (0.73) [0.73] & 1.63 (1.63) [1.63]& -- \\
0.00 & 0.25  & 3.25 & 0.78 (0.78) [0.79] & 1.77 (1.76) [1.76]& --  \\
0.00&  0.50&  3.50 & 0.84 (0.84) [0.84]& 1.90 (1.90) [1.90] &  --  \\
0.00&  0.75& 3.75 & 0.90 (0.89) [0.90]& 2.02 (2.03) [2.03] & -- \\
0.00&  1.00& 4.00 & 0.95 (0.95) [0.95]& 2.17 (2.17) [2.17] & -- \\
\hline

0.25 & 0.00  & 3.25 & 0.73 (0.77) [0.68]& 1.54 (1.45) [1.99] & 2.98 (2.83) [1.99] \\
0.25 & 0.25  & 3.50 &0.78 (0.80) [0.76] & 1.71 (1.67) [2.09]& 2.97 (3.00) [2.09]\\
0.25 & 0.50  & 3.75 &0.88 (0.85) [0.82] & 1.78 (1.84) [2.22]& 2.95 (3.03) [2.22]\\
0.25 & 0.75&  4.00 & 0.90 (0.91) [0.87]& 1.99 (2.00) [2.35] & 3.12 (3.07) [2.35] \\
0.25 & 1.00  & 4.25 & 0.95 (0.96) [0.92] & 2.19 (2.16) [2.49] & 3.10 (3.13) [2.49]\\
\hline

0.50 & 0.00& 3.50 & 0.72 (0.78) [0.66] & 1.42 (1.29) [2.30]& 2.92 (2.79) [2.30]\\
0.50 & 0.25& 3.75 & 0.83 (0.80) [0.72]& 1.53 (1.55) [2.42] & 2.88 (2.96)  [2.42]\\
0.50 & 0.50& 4.00 & 0.85 (0.86)  [0.79] & 1.77 (1.77) [2.54] & 3.03 (3.00) [2.54]\\
0.50 & 0.75 & 4.25 & 0.90 (0.91) [0.85] & 1.99 (1.96) [2.65] & 3.12 (3.06) [2.65] \\
0.50 &1.00  & 4.50 & 0.96 (0.97) [0.91] & 2.18 (2.14) [2.76] & 3.12 (3.013) [2.76]\\
\hline

0.75 & 0.00  &3.75 &  0.71 (0.77) [0.63] & 1.18 (1.00) [2.63]& 2.88 (2.75) [2.63] \\
0.75 & 0.25 & 4.00 &  0.78 (0.79) [0.71] & 1.47 (1.37) [2.71] & 2.91 (2.92) [2.71]\\
0.75 & 0.50& 4.25 & 0.85 (0.85) [0.80] & 1.76 (1.67) [2.78] & 2.99 (2.98) [2.78]\\
0.75&  0.75& 4.50 & 0.90 (0.92) [0.87] & 2.04 (1.90) [2.87] & 3.05 (3.05)  [2.87] \\
0.75&1.00 &  4.75 & 0.97 (0.98) [0.93] & 2.20 (2.10) [2.97] & 3.12 (3.12) [2.97]\\
\hline

1.00 & 0.00  & 4.00 & 0.68 (0.68) [0.68]& --  & 2.82 (2.82) [2.82]\\
1.00 & 0.25  & 4.25 &  0.75 (0.76) [0.76]& -- & 2.87 (2.87) [2.87] \\
1.00 & 0.50 &  4.50 & 0.83 (0.83) [0.83]& --  & 2.98 (2.96)  [2.96] \\
1.00 & 0.75 & 4.75 & 0.91 (0.91) [0.91]& --  & 3.09 (3.04) [3.04]\\
1.00 & 1.00 & 5.00 & 0.98 (0.98) [0.98] & --  & 3.13 (3.13) [3.13]\\
\noalign {\smallskip} \hline
\end{tabular}
\end{table*}

Up to now we have discussed the spin moments resulting for the CPA
approach. In table \ref{table1} we have gathered the total and
spin magnetic moments from all three SC, CPA and VCA calculations
as a function. of the concentrations $x$ and $y$. For the SC we
have scaled the moment to a unit cell of 4 atoms in order to
compare them with the other two methods although for the
calculations we have used a unit cell of 8 or 16 atoms depending
on the values of the concentrations. Our first remark concerns the
total spin magnetic moments in the third column. We have included
just one value since all three methods predict for all compounds
the half-metallic character and thus a total spin moment in
agreement with the Slater-Pauling behavior. The most important
feature of this table is the atom-resolved spin moments within the
supercell calculations with respect to the CPA approach. SC takes
into account the local-effects (short-range interactions) contrary
to CPA which is a mean-field theory taking into account only the
long-range order. The atom-resolved spin magnetic moments are
practical the same for all three transition metal elements, Co, Cr
and Mn, irrespectively of the $x$ and $y$ concentrations within
both CPA and SC. Thus for the full-Heusler alloys under study the
short-range interactions have a minimal influence on the
electronic structure of these alloys and these alloys can be
accurately described by mean-field theories like CPA. Although
this result seems astonishing, mainly due to our experience from
other alloys, we should not forget that Heusler alloys are
close-packed systems of very high symmetry in all directions.
Their properties are largely governed by symmetry arguments (like
the origin of the gap or the fixed number of minority occupied
states \cite{GalaFull}) and substitution of an atom by one of a
neighboring chemical element only marginally affects the
properties of these compounds. This argument is also supported by
the VCA calculations. The substitution of the crystal by a virtual
one with atoms of fractional electronic charge gives the same
total spin moment and almost identical Co spin moments with
respect to both SC and the more sophisticated mean-field theory of
CPA. Thus Co atoms are not really sensitive to their local
environment as long as the symmetry is not broken and each Co atom
has four low-valence transition metal atoms and four $sp$ atoms as
first neighbors. In the table we include also the spin moments of
the pseudoatoms within VCA although they have no physical meaning.

\begin{figure}
\centering
\includegraphics[width=\linewidth]{fig6.eps}
\caption{Total density of states (DOS) in states/eV for two
different compounds using the  CPA  method and a smearing of
10$^{-4}$ Hartree (sp1) and a smearing of 10$^{-3}$ Hartree
(sp10). The Fermi level has been chosen to be the zero of the
energy axis.} \label{fig6}
\end{figure}

\begin{figure*}
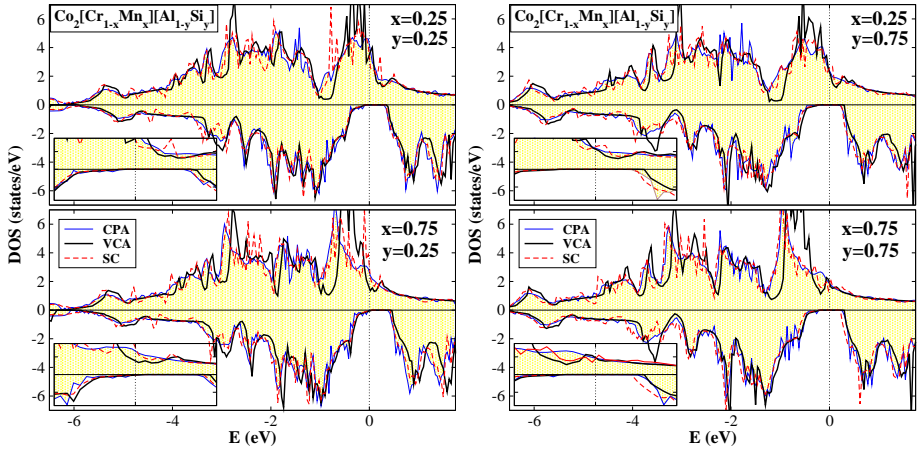

\centering
\includegraphics[scale=0.45]{fig7a.eps}
\includegraphics[scale=0.45]{fig7b.eps}
\caption{Total density of states (DOS) in states/eV for four
different compounds using all three CPA, VCA and SC methods.
Notice that the DOS for the SC calculations has been scaled to the
elementary unit cell of 4 atoms although the unit cell contains
either 8 or 16 atoms.} \label{fig7}
\end{figure*}

\section{Density of states \label{sec4}}

In the previous section we have discussed the half-metallicity of
the quinternary full-Heusler alloys in terms of the total spin
magnetic moments. In this section we will concentrate on the
density of states (DOS). This is important since we will use the
different DOS's to compute in detail the width of the gap, the
position of the Fermi level and the majority-spin DOS at the Femi
level. Thus we should be able to define accurately the edges of
the minority spin-gap. In VCA and SC calculations the gap is
clearly defined but problems arise within CPA. When we compute the
DOS within CPA we have to use a smearing for the DOS. Usually in
previous publications with the same method (see references
\cite{PSS-RRL} and \cite{SSC} for example) we have used a smearing
of 10$^{-3}$ Hartree but as shown in figure \ref{fig6} for two
different compounds this yields a region of very low DOS instead
of a real gap due to tails from both gap-edges which overlap. Thus
such a large smearing is  not suitable for the gap-related
properties. If we use a smearing of 10$^{-4}$ Hartree the
calculation of DOS is much more tedious but as shown in figure
\ref{fig6} we clearly get a region zero DOS and we are able to
define the edges of the gap. Thus in all CPA calculations which we
will present in the following we have used a broadening of
10$^{-4}$ Hartree.

In figure \ref{fig7} we have plotted the total DOS for four
different compounds taking all possible combinations of $x$ or
$y$= 0.25 or 0.75. All three methods CPA, VCA and SC produce
similar total DOS (note that for SC we have scaled the DOS to a
unit cell of four atoms). Especially CPA and SC give the same
positions for the bands of all atoms and the DOS's almost
coincide. VCA gives bands with the same mass center as CPA and SC
but with different shape.  This difference in shape with respect
to CPA is expected since VCA is an oversimplified mean-field
approach. If we concentrate on the region around the gap which we
have blown up in the insets it is obvious that all three methods
give similar description and similar values for the width of the
gap and the position of the Fermi level. This will be analyzed in
detail in the next section. These results support our statement in
the previous section that short-range interactions are not
important for the description of the magnetic and electronic
properties of the half-metallic Heusler alloys when the disordered
site contains atoms of neighboring chemical elements.

\begin{figure}
\centering
\includegraphics[width=\linewidth]{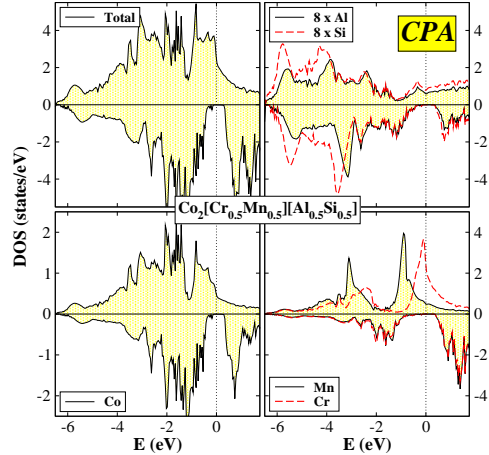}
\caption{Total and atom resolved DOS of the
Co$_2$[Cr$_{0.5}$Mn$_{0.5}$][Al$_{0.5}$Si$_{0.5}$] alloy using the
CPA method. The Al and Si DOS has been multiplied by a factor of
8. All atom-resolved DOS have been scaled to one atom}
 \label{fig8}
\end{figure}

\begin{figure}
\centering
\includegraphics[width=\linewidth]{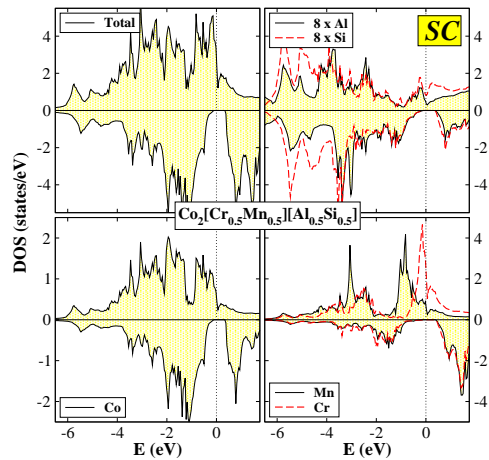}
\caption{Same as Fig. \ref{fig8} using the supercell construction.
The total DOS has been scale as before to one elementary unit cell
of four atoms.}
 \label{fig9}
\end{figure}

Finally in figures \ref{fig8} and \ref{fig9} we plot the total and
atom-resolved DOS for the
Co$_2$[Cr$_{0.5}$Mn$_{0.5}$][Al$_{0.5}$Si$_{0.5}$] compounds
within the CPA and SC approaches respectively. All DOS's have been
scaled to one atom and we have multiplied the DOS of the $sp$
atoms by a factor of 8 to make it visible. We do not present the
VCA results since atomic DOS's lack any physical meaning within
this approach. CPA due to the very small broadening used for the
DOS's gives a more spiky DOS with respect to SC. The Co DOS
produced by both methods is identical. In the majority spin DOS
the deep in the DOS around -1.5 eV separates the occupied majority
bonding d-states from the occupied antibonding majority
$d$-states. As expected for the Cr and Mn atoms the DOS present
several small differences but overall the bands present the same
shape. CPA is a mean field theory and thus washes out any small
picks in the DOS and for example we cannot distinguish between the
$t_{2g}$ and $e_g$ majority states of Cr around the Fermi level.
Contrary to CPA, SC carries also the short-range information and
the Fermi level separates the $t_{2g}$ and $e_g$ majority states
of Cr giving a band with a double peak. But overall the bands are
located at the same energy within both CPA and SC and have the
same width resulting in similar spin moments. CPA and SC give
somehow more important differences in the DOS of the Al and Si
atoms. But this DOS is one order of magnitude smaller than the DOS
of the transition-metal atoms since it is mainly made up from $p$
states which are spread over a wide energy range (the $s$ states
are located at around -9 eV and are not shown here). Thus these
differences do not affect the electronic description given by the
two methods and one can safely state that short-range interactions
are not important for these alloys.

\section{Gap-related properties\label{sec5}}

\begin{figure*}
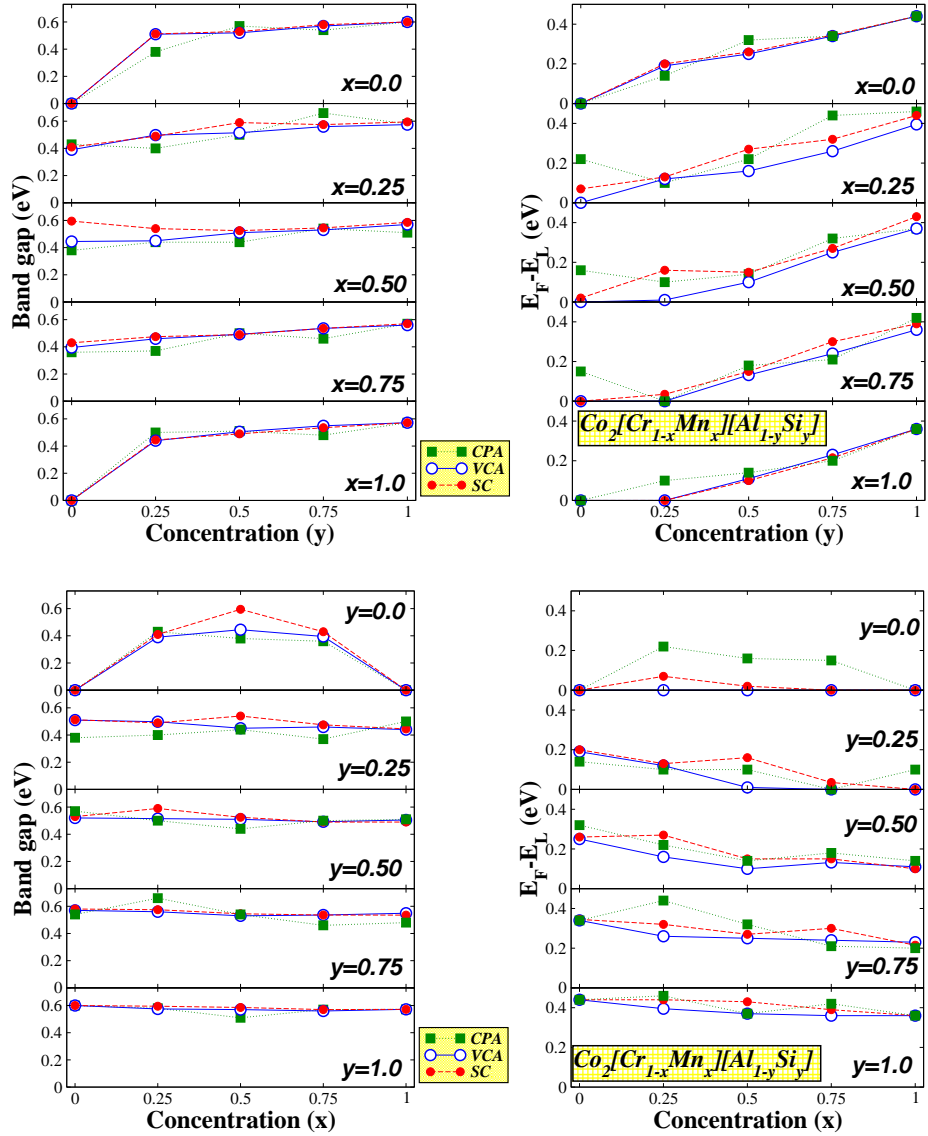

\centering
\includegraphics[scale=0.45]{./fig10a.eps}
\includegraphics[scale=0.45]{./fig10b.eps}
\vskip 0.5cm
\includegraphics[scale=0.45]{./fig10c.eps}
\includegraphics[scale=0.45]{./fig10d.eps}
\caption{Upper panel : width of the band gap (left panel) and
distance of Fermi level, $E_\mathrm{F}$, from the left edge of the
gap, $E_\mathrm{L}$, (right panel) as a function of the
concentration $y$ keeping $x$ constant using all three CPA, VCA
and SC methods. Lower panel : same as upper panel as a function of
$x$ keeping $y$ constant. Note that for the band-gap width the
scale of the vertical axis is different than for the position of
the Fermi level.} \label{fig10}
\end{figure*}

In the last part of our study, we present the properties related
to the gap. In figure \ref{fig10} we present the width of the gap
as a function of the concentration $y$ for constant values of $x$
in the upper left panel within all three approaches CPA, VCA and
SC and in the upper right panel the position of the Fermi level
with respect to the left edge of the gap. In the lower panel we
present the same information as a function of the concentration
$x$ of the transition metal atoms keeping the $y$ constant. Note
that for the width of the gap the vertical axis  goes up to 0.7 eV
while for the position of the Fermi level up to 0.5 eV. In table
\ref{table2} we present the majority density of states exactly at
the Fermi level. These results show the possibility of engineering
the properties related to the gap just by changing the
concentration in the low-transition metal and the $sp$ atoms. For
realistic applications we need three conditions (i) we should have
a quite large gap in order to have stable half-metallicity, (ii)
the Fermi level should be as close as possible to the center of
the gap for the half-metallicity to be robust since impurities and
defects induce states at the edges of the gap \cite{Orgassa}, and
(iii) the majority DOS at the Fermi level should be also high in
order to produce significant spin-polarized current in real
experiments.

The first remark concerning figure \ref{fig10} is that all three
methods give almost identical results for the width of the
minority-spin gap. This enhances even further our argument in
section \ref{sec3} that short-range interaction taken into account
in the supercell calculations are not significant for the
full-Heusler alloys. In the case of the alloys rich in Si
($y$=0.5, 0.75 and 1.0) the width of the gap is constant and
around 0.6 eV, while for the compounds rich in Al ($y$=0.0 and
0.25) the gap is around 0.4-0.5 eV and in the case of Co$_2$CrAl
and Co$_2$MnAl alloys it almost vanishes. Thus for realistic
applications compounds rich in silicon should be preferable since
the gap-width should be large and more robust. From the low-left
panel we can safely conclude with the exception of the compounds
containing exclusively Al ($y=0$) that the width of the gap is
independent of the relative concentration of the Cr and Mn atoms
($1-x$ and $x$ respectively) as we keep constant the
concentrations of the Al and Si atoms. The gap is defined by the
relative position of the non-bonding coccupied triple-degenerated
$t_{1u}$ and unoccupied double-degenerated $e_u$ states which are
exclusively located at the Co sites as shown in reference
\cite{GalaFull}. The position of these states does not depend on
the exact energy position of the Cr and Mn $d$ orbitals and thus
the width of the gap does not depend on the relative concentration
of these atoms.

Now we will discuss the behavior of the Fermi level. The Fermi
level contrary to the width of the gap is determined mainly by the
energy extension of the majority $p$ states of the $sp$ atoms as
it was discussed in reference \cite{GalaFull}. This is clearly
seen in the upper right panel of figure \ref{fig10} where, when we
keep $x$ constant, the Fermi level is pushed higher in energy as
we increase the concentration of the Si atoms with respect to the
Al ones.The Fermi level is located almost at the left edge of the
gap for the alloys containing only Al ($y$=0) and has a distance
of $\sim$0.4 eV from the left edge of the gap for the alloys
containing only Si ($y$=1). Since the most interesting case are
the alloys rich in Si with a gap around $0.6$ eV, the ideal case
is that the Fermi level is located at the middle of the gap with a
distance of about 0.3 eV from both edges of the gap. This is the
case when $y=0.75$ and thus at the Z site 25\%\ of the atoms are
of Al chemical type and 75\%\ of the Si chemical type. When
$y=0.5$ and the populations of Al and Si atoms are equal the Fermi
level is located at around 0.2 eV from the left edge of the gap
and thus half-metallicity should be more affected by disorder,
defects and impurities.

\begin{table}
\centering \caption{Calculated spin-up (majority) density of
states in states/eV at the Fermi level for the
Co$_2$[Cr$_{1-x}$Mn$_x$][Al$_{1-y}$Si$_y$] as a function of the
$x$ and $y$ concentrations with three different methods (i) the
coherent potential approximation (CPA), (ii) the virtual crystal
approximation (VCA, and (iii) using supercell calculations (SC).
Note that for the SC case the total DOS has been scaled as before
to an elementary unit cell of four atoms instead of 8 or 16 atoms
of the unit cell. \label{table2}}
\begin{tabular}{r|r|r|r|r}\hline\noalign{\smallskip}
$x$ & $y$ & CPA & VCA & SC \\ \hline
 0.00 & 0.00  & 4.861 &    4.861  &  4.861 \\
 0.00 & 0.25  & 4.552&4.494& 5.102\\
  0.00&  0.50&   3.673&     3.196 &   3.660\\
   0.00&  0.75& 2.967&2.663 &3.170\\
    0.00&  1.00&   2.318 &    2.318 &   2.318\\ \hline
0.25 & 0.00  & 3.030   &  4.490  &  5.189 \\
0.25 & 0.25  & 3.496& 4.201 &3.339 \\
0.25 & 0.50  & 3.553 &  3.110& 2.488\\
0.25 & 0.75& 2.271 &2.381 &2.544 \\
0.25 & 1.00  & 1.332 &    2.092 &2.319\\ \hline

0.50 & 0.00& 4.746 &3.861 &   4.528 \\
0.50 & 0.25&   3.843 &  3.581 &4.240\\
 0.50 & 0.50& 1.906&  2.477&2.804\\
 0.50 & 0.75 & 1.270 &2.088& 2.016\\
0.50 &1.00  & 1.904 &    1.861& 1.974\\ \hline
 0.75 & 0.00  & 3.670&2.910  &  4.027 \\
 0.75 & 0.25 &  1.543 &    2.970 &3.051\\
 0.75 & 0.50&1.512 & 2.210  &2.083\\
  0.75&  0.75&   1.896& 1.851  &  1.787\\
 0.75&1.00 &  1.707 &    1.542  &  1.461\\ \hline
1.00 & 0.00  & 1.510   &  1.510 &   1.510 \\
1.00 & 0.25  & 2.042 &2.025 &1.876 \\
1.00 & 0.50 &  1.582 & 1.556 & 1.610 \\
1.00 & 0.75 &1.451 &1.203& 1.491\\
 1.00 & 1.00 &  1.222 & 1.222&1.222\\
\noalign {\smallskip} \hline
\end{tabular}
\end{table}

Finally, we shall focus on the majority density of states
presented in table \ref{table2}. All three methods give comparable
results but now the SC approach gives systematically a higher
value than CPA or VCA with few exceptions. The DOS at the Fermi
level is a small detail of the total DOS and the effect of the
short-range interactions is larger than when we compare the DOS
over the whole energy range or the spin moments where these small
details are averaged and do not manifest their presence. As we
have discussed in section \ref{sec4} the DOS at the Fermi level is
higher for the compounds rich in Cr, since for the Cr atoms the
Fermi level crosses the majority $d$ electrons while Mn atoms have
practically all their $d$ majority states occupied and the
majority DOS at the Fermi level for the compounds containing only
Mn ($x=1.0$) is one third the value for the compounds containing
only Cr ($x=0.0$). A second remark concerns the behavior of the
majority DOS at the Fermi level with the concentration $y$ of the
sp atoms when we keep $x$ constant. As we increase the population
in Si the extra charge pushes the majority states lower in energy
and thus the DOS at the Fermi level deceases considerably and for
the compounds containing only Si ($y=1.0$) the DOS at $E_F$ is
about 40-50\%\ of the value for the Al alloys ($y=0.0$). As
discussed in the previous paragraphs for realistic applications we
need compounds rich in Si which combine large a band-gap width
with a Fermi level near the middle of the gap. From the discussion
in this paragraph it is evident that we also need our compound to
be rich in Cr in order to ensure a high value of DOS at the Fermi
level.

\section{Summary and conclusions \label{sec6}}

We have reviewed the electronic and magnetic properties of the
quinternary full Heusler alloys of the type
Co$_2$[Cr$_{1-x}$Mn$_x$][Al$_{1-y}$Si$y$]. For our study we have
employed three different approaches the coherent potential
approximation (CPA), the virtual crystal approximation (VCA) and
supercell calculations (SC) to have information also for the short
range interactions. All three methods gave similar results and the
local environment manifested itself only for small details of the
density of states.

All three approaches predicted that the alloys under study are
half-metals for all $x$ an $y$ concentrations and their total spin
moments follow the so-called Slater-Pauling behavior : the total
spin magnetic moment in the unit cell in $\mu_B$ equals the number
of valence electrons in the unit cell minus 24. The spin moment of
the transition metal atoms were found to be almost insensible to
the relative concentrations of Cr and Mn elements while the $sp$
atoms carried very small antiparallel spin moments.

All three CPA, VCA and SC approaches yielded similar results for
the properties related to the minority-spin band-gap. The width of
the gap is determined by states exclusively localized at the Co
sites and is insensitive to the Cr and Mn concentrations and is
larger for the compounds rich in Si.  The Fermi level is
positioned at the left edge of the gap for the alloys containing
only Al and as we increase the concentration in Si it is pushed
higher in energy. The majority-spin density of states at the Fermi
level takes larger values for the compounds rich in Cr and it
drops as we increase the Si concentration.

We have shown the possibility to engineer the properties of the
half-metallic Heusler alloys by changing the concentration of the
low-valent transition metal and $sp$ atoms in a continuous way. We
conclude that for realistic applications ideal are the compounds
rich in Si and Cr since they combine large energy gaps (around 0.6
eV), robust half-metallicity with respect to defects (the Fermi
level is located near the middle of the gap) and high values of
the majority-spin density of states around the Fermi level. Since
such alloys are realized already experimentally in the form of
thin films and multilayers and are incorporated in
magnetoelectronic devices, like spin-valves or magnetic tunnel
junctions, we expect our results to be of interest for the
community of experimentalists in the field of spintronics and to
stimulate further experimental interest on these compounds.

\end{document}